\documentclass{sig-alternate-05-2015}

\usepackage{graphicx}
\usepackage{subcaption}

\begin{document}
\setcopyright{acmcopyright}

\CopyrightYear{2017} 
\setcopyright{acmcopyright}
\conferenceinfo{FPGA '17,}{February 22-24, 2017, Monterey, CA, USA}
\isbn{978-1-4503-4354-1/17/02}\acmPrice{\$15.00}
\doi{http://dx.doi.org/10.1145/3020078.3021738}

%

\title{An {OpenCL\textsuperscript{\texttrademark}} Deep Learning Accelerator on Arria 10}
%
%
%
%
%

\numberofauthors{1} 
%
\author{
%
%
 Utku Aydonat, Shane O'Connell, Davor Capalija, Andrew C. Ling, Gordon R. Chiu\\
       \affaddr{Intel Corporation}\\
       \affaddr{Toronto, Canada}\\
       \email{utku.aydonat|shane.oconnell|davor.capalija|andrew.ling|gordon.chiu@intel.com}
}

\maketitle
\begin{abstract}
Convolutional neural nets (CNNs) have become a practical means to
perform vision tasks, particularly in the area of image classification. FPGAs
are well known to be able to perform convolutions efficiently, 
however, most recent efforts to run CNNs on FPGAs have shown limited
advantages over other devices such as GPUs.
Previous approaches on FPGAs have often been memory bound
due to the limited external memory bandwidth on the FPGA device. We show a
novel architecture written in OpenCL\texttrademark, which we refer to as a Deep Learning Accelerator
(DLA), that
maximizes data reuse and minimizes external memory bandwidth.
Furthermore, we show how we can use the Winograd transform to significantly
boost the performance of the FPGA.
As a result, when running our DLA on Intel's Arria 10 device we can achieve a performance of 1020{\small img/s}, or 23{\small img/s}/W when running the AlexNet CNN benchmark. This comes to 1382 GFLOPs and is 10x faster with 8.4x more GFLOPS and 5.8x better efficiency
than the state-of-the-art on FPGAs. Additionally, 23{ \small img/s}/W is competitive against 
the \emph{best publicly known} implementation of AlexNet on nVidia's TitanX GPU.
\end{abstract}


\terms{FPGA, Convolution, Neural Network}
\vspace{-0.1in}


\vspace{-0.1in}
\section{Introduction}
Convolutional neural nets (CNNs) have become widely adopted in various
computer vision applications including driver assist and image
classification. More recently, FPGAs have shown
promise in efficiently implementing CNNs~\cite{Zhang2015,Suda2016,Qiu2016,Zhang2016,Microsoft,Cadambi2010,Peemen2013,Sankaradas2009,Farabet2009}.
Unfortunately, the vast majority of FPGA implementations of CNNs
have only implemented the convolutional layers limiting the
benefit of the approach since other layers may quickly become the
bottleneck of the neural net~\cite{Zhang2016}.
There has been work in
implementing all the layers on the FPGA~\cite{Zhang2016,Suda2016,Qiu2016}, however, when compared to some of the best results publicly known for GPUs~\cite{ConvNet,NvidiaTitanX},
FPGA performance have fallen short significantly.

One of the reasons that FPGAs have not been able to achieve good
performance against GPUs is due to their limited external memory bandwidth.
CNNs are often solved using matrix-multiplication based approaches, which require
large amounts of data to be moved between the compute units and
external memory~\cite{Suda2016}. Additionally, previous FPGA architectures for CNNs have not
been able to take advantage of the peak operations of the device
leading to low performance~\cite{Zhang2015,Suda2016,Qiu2016,Zhang2016}.

To address the problems above, we introduce a novel architecture
described in OpenCL and provide the following contributions:
{\small
\begin{itemize}
	\item A methodology to minimize bandwidth of convolutional and  fully-connected layers by caching all intermediate feature-maps on-chip in \emph{stream-buffers}.
		In conjunction with batching images during fully-connected layers, which is similar to what is used in~\cite{Zhang2016}, we are able to reduce
		the external bandwidth requirements by an order-of-magnitude  for both the convolutional and fully-connected layers.
\item A design space exploration methodology that leverages
    analytical models for resource usage and throughput and is able to find
	the optimal architecture configuration, for a specific FPGA device and CNN, to get maximum throughput.
\item An approach that leverages the Winograd transformation to reduce the
    multiply-accumulate operations of the convolutions~\cite{Winograd}.
\end{itemize}
}

Due to the contributions above we are able to implement all layers of AlexNet~\cite{AlexNet}
on Intel's Arria 10 FPGA and achieve over 10x better throughput and
8.4x more GFLOPS than the state-of-the-art
FPGA implementation of AlexNet~\cite{Zhang2016}. Furthermore, we show that, to the best of our knowledge, this is the first FPGA implementation whose performance per watt
is competitive against the same generation \emph{highly-optimized} TitanX GPU results~\cite{ConvNet,NvidiaTitanX,M4Nvidia}.

The rest of the paper is organized as follows.
Section~\ref{sec:background} has background on CNNs and related
work. Section~\ref{sec:cnn_architecture} describes the DLA architecture.
Section~\ref{sec:performance_modeling} describes our analytical model
for design space exploration. Finally,
Sections~\ref{sec:experimental_eval} and~\ref{sec:results} describe our
results.

\section{Background}
\label{sec:background}

Deep neural networks are machine learning algorithms that are inspired by the structure and function of the human brain. They consist of several interconnected artificial neurons that are modeled after the neurons of the human nervous system. An artificial neuron accepts numerical input from other neurons, and produces an output. For DNNs, the output is computed as a dot-product of its inputs and its unique set of learnable weights. Subsequently, a non-linear \textit{activation function} (e.g. $\tanh$, ReLU, sigmoid) is applied to the dot-product result. This output is then used as input by other neurons. Neural networks have been used to solve many complex problems to which robust solutions cannot be designed by hand such as image recognition, handwritten text, gesture, and speech recognition; game-playing and decision making (e.g. AlphaGo); face identification; and object detection.

\vspace{0.1in}
\subsection{Convolutional Neural Networks}
\label{sec:convolutional_neural_networks}

Convolutional neural networks (CNNs) are neural networks that excel in classifying images and videos.
They have garnered a considerable amount of attention in recent years due their ability to achieve state-of-the-art results in image recognition and object detection. CNNs are neural nets that consist primarily of \textit{convolution} layers in which each neuron is connected only to a small, nearby region of neurons in the previous layer. This local connectivity is intentionally designed into the network topology with the goal of exploiting the local correlation in the input data. This connectivity restriction, together with the additional property that groups of neurons within one convolution layer also share learnable weights, allows the outputs of neurons in the layer to be computed using 3-dimensional convolution.
\\
\\
Although a CNN can be described from a neuronal perspective, it is more instructive, for the discussion that follows, to view it as a directed graph of \textit{computational layers}. Each node represents a layer that accepts one or more $n$-dimensional arrays as input, performs some computation, and produces one or more $n$-dimensional arrays as output. The edges of the graph represent the producer-consumer relationships between the layers of the network. The data arrays that layers within the network consume and produce are often referred to as \textit{feature maps}.
\\
\\
In AlexNet, a convolution layer accepts a 3-dimensional array with depth $C$, height $H$, and width $W$ as input, and produces a 3-dimensional array with depth $K$, height $P$, and width $Q$. The output feature map is computed by convolving the input feature map with $K$ \textit{filters}, and applying an activation function element-wise to the result. Each filter is also a 3-dimensional array with depth $C$, height $R$, and width $S$ which consists of learnable weights. The convolution of the input feature map with one filter produces one 2-dimensional array referred to as a \textit{channel} or \textit{plane} of the output feature map. The entire output feature map is obtained by concatenating depth-wise the $K$ channels produced by convolving each of the $K$ filters with the input feature map. An illustration of the images is shown in Figure~\ref{fig:cnn_overview}.

\subsection{AlexNet}
\label{sec:alexnet}

AlexNet~\cite{AlexNet} consists of the following layers:

{ \small
\begin{itemize}
    \item \textbf{Convolution} - The previous section describes the
	functionality of convolution layers.
    In AlexNet, all convolution layers use the ReLU or ramp function $f(x) = \max(0, x)$ as their activation function. In addition, each convolution layer also has $K$ scalar bias terms that are added to corresponding output feature map channels before applying the ReLU function.
\item \textbf{Cross-channel local response normalization} - A normalization layer scales each element in its input feature map by a factor that is a function of the elements at the same location in adjacent channels as the element being normalized. The dimensions of the output and input feature maps are identical.
\item \textbf{Max pooling} -  A max pooling layer strides a two-dimensional window across each channel of the input feature map and propagates the element of maximum value in the window through to the output feature map. Compared to the input feature map, the output feature map has the same depth, smaller height, and smaller width.
\item \textbf{Fully-connected(dense)} - A fully connected layer is a convolution layer in which $H = R$ and $C = W = S = 1$ (which in turn implies that $P = Q = 1$). That is, the height and width of each filter is equal to the height and width of the input feature map. Described from a neuronal perspective, a fully-connected layer is one in which each neuron is connected to every neuron in the previous layer (hence the name fully-connected). Since the input feature map and each filter have the same dimensions, no striding occurs when computing the output feature map. As a result, it is more convenient to think of the output as a matrix-vector product $\mathbf{v_o} = \mathbf{Wv_i}$ where $\mathbf{v_i}$ is a flattened version of the input feature map containing $n_i = C \times H \times W$ elements, $\mathbf{W}$ is a $n_o = K$ by $n_i$ matrix in which row $k$ is a flattened version of the $k^{th}$ filter, and $\mathbf{v_o}$ is the output feature map. It is possible to process a batch of $b$ different input feature maps from $b$ different images at once by replacing $\mathbf{v_i}$ with an $n_i$ by $b$ matrix $\mathbf{V_i}$ in which column $k$ is the flattened input feature map corresponding to the $k^{th}$ image in the batch. The aforementioned equation then becomes $\mathbf{V_o} = \mathbf{WV_i}$, where $\mathbf{V_o}$ is an $n_o$ by $b$ matrix in which column $k$ is the flattened output feature map corresponding to the $k^{th}$ image in the batch. This method of processing multiple images at once in a fully-connected layer will feature prominently in the upcoming discussion.
\item \textbf{Softmax} - A softmax layer normalizes the values in the input feature map by applying the softmax function to it. Consequently, the sum of the elements in the output feature map is unity.
\end{itemize}

}
At a high level, AlexNet consists of five convolution layers, followed by three fully-connected layers, and a softmax layer. There is a normalization layer after the each of the first two convolution layers. Finally, there is a max-pooling layer after the two aforementioned normalization layers, and between the last convolution layer and the first fully-connected layer. The final softmax layer outputs a 1000-element vector containing probabilities that the input image belongs to each of the 1000 possible classes in the ImageNet Large Scale Visual Recognition Competition (ILSVRC~\cite{ILSVRC}). More details regarding the structure and function of AlexNet can be found in~\cite{AlexNet}.

\subsection{Related Work}
\label{sec:rel_work}
FPGAs have been shown to be a practical means to solve CNNs~\cite{Zhang2015,Suda2016,Qiu2016,Zhang2016,Microsoft,Cadambi2010,Peemen2013,Sankaradas2009,Farabet2009}.
In~\cite{Suda2016}, the authors use a matrix-multiply approach to solve
both convolutional and fully-connected layers which is similar
to GPU and CPU approaches that convert 3D convolutions into 2D
matrix-multiplications. Written in OpenCL, they
are able to run all layers on the FPGA but unfortunately end up being
severely external memory bound such that the average GOPs they achieve
is relatively low.
To solve the memory bottleneck, in~\cite{Qiu2016} the authors introduce a singular value decomposition approach to significantly reduce the data required, and hence memory bandwidth, of the fully connected layers. They empirically show that this has approximately 1\% impact on the overall accuracy of the neural network when applied to image classification.

Conversely, the work in~\cite{Zhang2015,Zhang2016} use a roofline model that
allows users to maximize compute resources on the FPGA given the memory
bandwidth constraints.
In~\cite{Zhang2016}, the authors describe Caffeine which is a runtime
reconfigurable CNN FPGA accelerator. In Caffeine, the throughput is improved significantly over previous
approaches by 
creating a model to realistically reflect DDR
transfers and also provide a \emph{convolutional MM representation} where they are able to maximize data reuse of weight filters by batching input feature maps of the fully-connected layers. They show that they are able to improve the performance of
CNN on FPGAs by 3x, and they are 1.5x more energy efficient than the K40 GPU.  
Unfortunately, when compared to nVidia's last generation TitanX GPU~\cite{ConvNet,NvidiaTitanX} the
power efficiency of the FPGA is still 5.8x worse. Additionally, the
authors of~\cite{Zhang2016} show that the GOPs of each layer is relatively low where they are only able to achieve 14.7\%
of the GOPs of the KU060 device when running at 200MHz.

Our approach differs from the previous work as we
significantly reduce memory bandwidth without loss of accuracy by caching all feature-maps on-chip.
Additionally, we show that our architecture is \emph{compute-bound}, such that we can efficiently use
all the DSP resources, and ensure that they are occupied (i.e. doing
useful work) the majority of the time and leverage Winograd transforms to reduce the number of required operations. 
Finally, we show how we use a design space exploration methodology to find the optimal configuration of our architecture for a specific FPGA device and CNN. All of these factors lead to
a performance efficiency that is competitive against nVidia's TitanX GPU.

\subsection{Intel FPGA SDK for OpenCL}
\label{sec:opencl}
The Intel FPGA SDK for OpenCL allows users to program
FPGAs with OpenCL. OpenCL is an open parallel programming
language that is vendor agnostic and is supported by many
vendors~\cite{OpenCLKhronos}.

\begin{figure}
\centering
\includegraphics[scale=0.25]{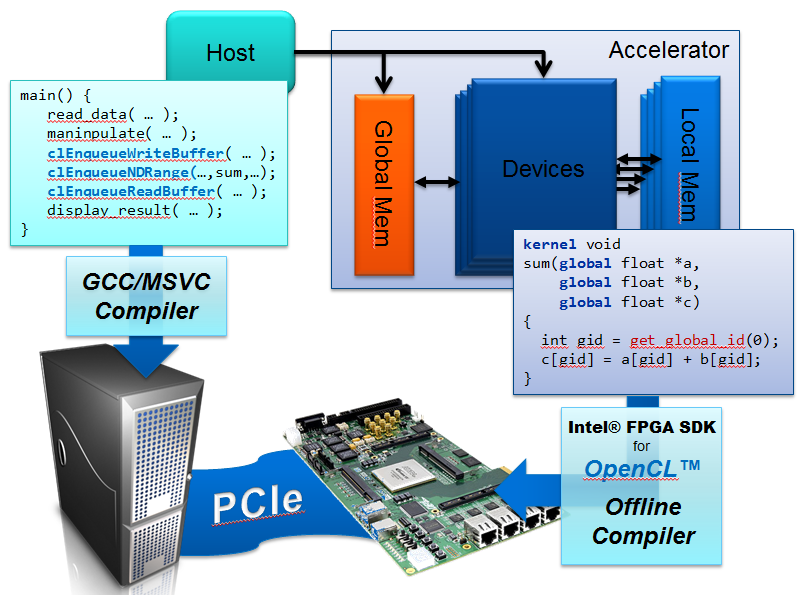}
\caption{Intel FPGA SDK for OpenCL Host-Device Setup and Flow.}
\label{fig:master_slave}.
\end{figure}

Currently, OpenCL uses a master-slave model where a master host device is
used to control all memory transfers and execution of the kernels. A
user is required to write a host program, which calls a 
predefined OpenCL API to control the accelerator device. On
the device side, the user writes OpenCL kernel functions that are
compiled to the accelerator. This model is illustrated in
Figure~\ref{fig:master_slave}.

\begin{figure}
\centering
\includegraphics[scale=0.5]{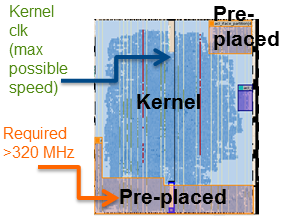}
\caption{OpenCL FPGA Platform on an Intel's Device.}
\label{fig:platform}
\end{figure}

One of the key challenges for using FPGAs is that they have
traditionally required a hardware design methodology. 
Because of the reconfigurable nature of FPGAs, timing-sensitive
components such as DDR memory controllers must be timing closed to ensure they work
correctly.
The Intel FPGA SDK for OpenCL avoids these problems by providing a
pre-generated platform for the OpenCL programmer. An illustration of the
platform on the FPGA device is shown in Figure~\ref{fig:platform}. As
illustrated, the platform has pre-placed components whose resources are
reserved for the platform, and cannot be used for the algorithmic
portion of the OpenCL kernel code.

Our DLA is written with OpenCL, where OpenCL kernels are used to define
the DLA architecture, and the host runtime is used to coordinate the
data transfers of the images with the kernel execution in an efficient
manner.

\section{DLA Architecture}
\label{sec:cnn_architecture}

Our Deep Learning Accelerator (DLA) implements \emph{all layers} of AlexNet on the FPGA and is defined using the Intel FPGA SDK for OpenCL.

\subsection{Design Goals}
\label{sec:design_goals}


Our DLA is targeted for high-performance.
In most CNN topologies, the total amount of floating-point computation is dominated by the convolution layers.
For instance, in AlexNet, convolutions are 92\% of the total floating point operations.
Hence, the DLA hardware is optimized to maximize the throughput of the convolution layers by exploiting parallelism in computations.
Each convolution layer consists of multiple nested loops that iterate over the dimensions of input features, filters, and output features.
As shown in~\cite{Zhang2015}, it is possible to choose different combinations of these loops to vectorize in order to speedup the convolution operations.
On the FPGA, vectorizing a loop means spatially distributing the computations of that loop across multiple DSP blocks that exist on the device.
For maximum performance, our DLA vectorizes the loops that provide sufficient parallelism such that as many DSPs as possible are used every cycle for useful computations.
Additionally, the DLA architecture ensures that the processing elements (PEs) are able
to solve both the convolutional and fully-connected layers without
sacrificing performance.

Our DLA is also aimed to be flexible and achieve good performance with other CNN topologies, besides AlexNet.
Hence, convolution loops are chosen for vectorization such that enough parallelism exists not just in AlexNet but in a wide range of CNN topologies.
Consequently, adapting our DLA for a different CNN topology will not require vectorizing different loops, but will just require changing the vectorization factors according to the dimensions of that topology. This is similar to what is claimed in~\cite{Zhang2016} and is not discussed in detail in this work.

\vspace{-0.05in}
\subsection{Convolution Layers}
\label{sec:two_dimensions_of_parallelism}

\begin{figure}
\centering
\includegraphics[scale=0.4]{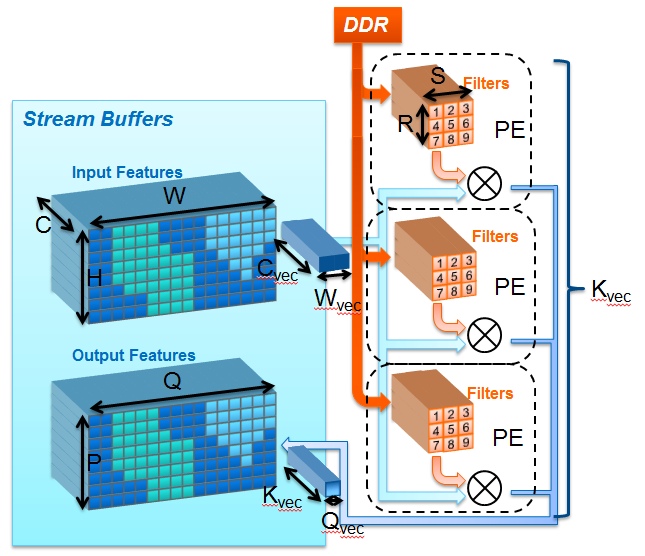}
\caption{The overview of Convolution Execution.}
\label{fig:cnn_overview}
\end{figure}

To improve throughput, parallelism is extracted from four dimensions of a convolution layer: output feature columns ($Q$), output feature maps ($K$), input feature maps ($C$), and input feature columns ($W$).
The vectorization factors for each of these dimensions are respectively referred to as $Q_{vec}$, $K_{vec}$, $C_{vec}$, and $W_{vec}$.
Each cycle, $Q_{vec}$ horizontal output features in $K_{vec}$ output feature maps are computed by convolving an input feature region $W_{vec}$ wide and $C_{vec}$ deep.
This is illustrated in Figure~\ref{fig:cnn_overview}, for $C_{vec}>1$,
$K_{vec}=3$, $W_{vec}=2$, and $Q_{vec}=1$.
The relationship between $W_{vec}$ and $Q_{vec}$ depends on the number of filter and feature pixels that are multiplied per output result. For example, in equation~\ref{equ:fir_filter}, for each output, a vector of three feature and filter pixels are used. In this case $W_{vec}=S_{vec}+Q_{vec}-1$, where $S_{vec}$ is the size of the filter vector (e.g. $S_{vec}=3$ in equation~\ref{equ:fir_filter}). If a larger $W_{vec}$ is desired, a larger $S_{vec}$ is required for each output computation.


Vectorizing the $W$ and $Q$ dimensions is also useful for the arithmetic optimizations which will be discussed in section~\ref{sec:arithmetic_optimizations}.
Because convolution layers usually process large number of input and output feature maps, enough parallelism can be extracted in these dimensions to use all the DSPs by breaking up the convolution operations into individual dot-products that are processed by PEs.


The PEs act as dot-product solvers for the features and filter weights.
Each PE receives the same input features, illustrated as a $1\times W_{vec} \times C_{vec}$ stick in Figure~\ref{fig:cnn_overview}, and convolves them with the filter weights, of size $1\times S_{vec}\times C_{vec}$, to produce a vector of $Q_{vec}$ output features for one output feature map. Hence, at any given time $K_{vec}$ PEs will be computing $K_{vec}$ different output feature-maps.
In other words, $K$ output feature maps are computed in $K$/$K_{vec}$ tiles.
Convolution layers are mapped onto the architecture in Figure~\ref{fig:cnn_overview} in a time-multiplexed fashion, i.e. layers are executed one at a time in succession.
This is possible because the sizes of $C_{vec}$, $K_{vec}$, $Q_{vec}$ and $W_{vec}$ can be independent of the image size, filter size, and number of input maps and output maps, thus solving different convolution layers simply requires different sequences of $1\times W_{vec}\times C_{vec}$ sticks of input feature maps and filter data to be read and sent to the PEs.

Our DLA takes advantage of the mega-bytes of on-chip storage available on the FPGA device by storing the features and the filter weights in on-chip RAMs.
The features are stored in a double buffer and data is broadcast to PEs in a daisy-chain fashion every cycle. The daisy-chain structure is formed by the PEs where each PE receives a stick of input feature data for processing, and also passes the data to an adjacent PE (PE daisy-chain arrangement illustrated in Figure~\ref{fig:overall_architecture}). This is much more efficient for placing the DLA on the FPGA since the FPGA is a 2D grid of logic.
The outputs of the PEs are stored back into the double buffer.
The filters are stored in caches inside the PEs.
The purpose of the on-chip storage is to avoid unnecessary external memory accesses because the amount of features and filter weights loaded every cycle depends on the vectorization parameters and can easily exceed the available external memory bandwidth.
On-chip storage allow the re-use of data by taking advantage of the temporal data locality.
More specifically, double buffers allow the re-use of the input feature maps because same input features are convolved with different filters to compute different output feature maps.
In addition, filter caches allow the re-use of the filter weights because same filter weights are convolved with different input features to compute each output feature map.

\vspace{-0.05in}
\subsection{Arithmetic Optimizations}
\label{sec:arithmetic_optimizations}
\begin{figure}
\centering
\includegraphics[scale=0.3]{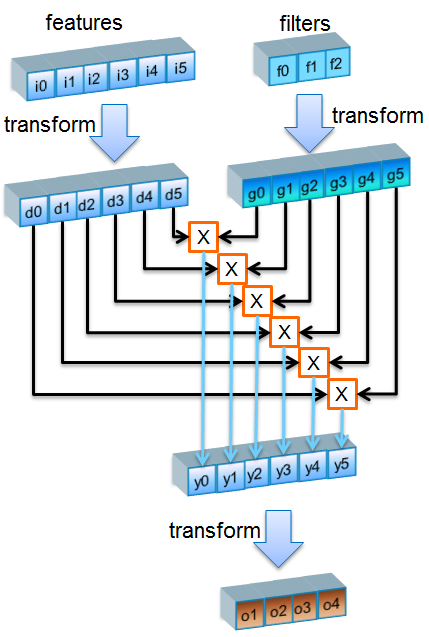}
\caption{Winograd Flow}
\label{fig:winograd_flow}
\end{figure}

In addition to providing parallelization, vectorizing on $W$ and $Q$ allows multiple multiply-accumulate operations to be simplified through Winograd transformations as described here~\cite{Winograd}.
Lavin et al.~\cite{DBLP:journals/corr/Lavin15b} showed that Winograd's minimal filter algorithms~\cite{Winograd} can be used to derive algorithms for CNNs.
These algorithms can be applied when the convolution stride is 1 and can reduce the arithmetic complexity, resulting in faster execution.
It has also been shown that the reduction in arithmetic complexity can exceed what can be achieved with other FFT-based methods for small filters.
The AlexNet topology uses small $3\times 3$ filters in most of its convolution layers, hence, can take advantage of the Winograd transformations.
Furthermore, because the current trend is towards deeper CNN topologies with small filters (e.g. GoogLeNet~\cite{DBLP:journals/corr/SzegedyLJSRAEVR14}) other CNN topologies can also take advantage of the Winograd transformations.

%

\begin{equation} \label{equ:fir_filter}
    \begin{split}
    o_0 &= (f_0,f_1,f_2)\cdot(i_0,i_1,i_2) \\
    o_1 &= (f_0,f_1,f_2)\cdot(i_1,i_2,i_3) \\
    o_2 &= (f_0,f_1,f_2)\cdot(i_2,i_3,i_4) \\
    o_3 &= (f_0,f_1,f_2)\cdot(i_3,i_4,i_5)
    \end{split}
\end{equation}
In our DLA, each PE generates four horizontal output pixels in parallel (i.e. $Q_{vec}=4$) where each output is formed by doing a dot-product between three filters and three inputs as shown in equation~\ref{equ:fir_filter}.
In standard convolutions, this requires 12 multiplications and additions every cycle which is shown in equation~\ref{equ:fir_filter} where $o_i$ is an output pixel, $f_i$ is a filter weight, and $i_i$ is an input pixel.
With the Winograd minimal filtering algorithms, we perform the four dot-products in equation~\ref{equ:fir_filter} with only six multiplications and additions using techniques described in~\cite{Winograd}, and denoted as $F(4,3)$. All Winograd arithmetic transformations are done on-chip and the flow is illustrated in Figure~\ref{fig:winograd_flow} which shows how we transform three filter coefficients and six feature inputs into six Winograd filters and six Winograd input features (i.e. $W_{vec}=6$). The six values are multiplied together to form six Winograd outputs, which then are transformed back to four output features.

\subsection{PEs}
\label{sec:pes}

\begin{figure}
\centering
\includegraphics[scale=0.3]{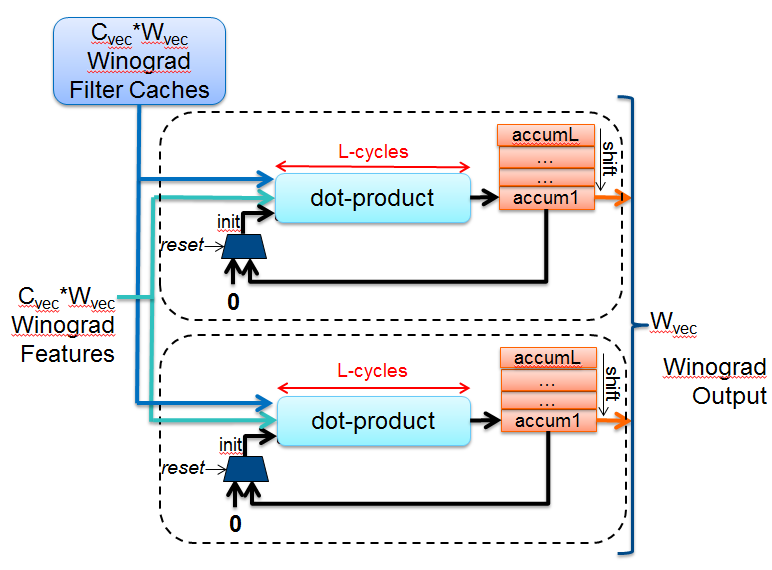}
\caption{The overview of a PE.}
\label{fig:pe}
\end{figure}

Figure~\ref{fig:pe} shows an overview of a single PE hardware. 
It consists of dot-product units, accumulators, and caches.

Each dot-product unit multiplies and accumulates the Winograd transformed input features and the filter weights.
The vector size of the dot-product unit is determined by the $C_{vec}$ parameter as shown in Figure~\ref{fig:pe}.
Each PE contains $W_{vec}$ of such dot-product units.
Hence, a sub-region of size $1\times W_{vec}\times C_{vec}$ is convolved every cycle.
Once the total input feature region is convolved, $Q_{vec}$ output features are completed.

Each dot-product unit takes as input $C_{vec}\times W_{vec}$ features, $C_{vec}\times W_{vec}$ transformed filter weights, and an {\em init} bus as shown in Figure~\ref{fig:pe}. To support Winograd,
we take $C_{vec}\times S_{vec}$ filters and convert them to $C_{vec}\times W_{vec}$ transformed filter weights.
The {\em init} bus is set to zero when {\em reset} is set, which represents the start of a new output feature computation.
If {\em reset} is not set, then {\em init} is set to the current accumulator value so that the accumulator is incremented by the dot-product result.
If the {\em done} signal is set, the dot-product result is sent out.
This happens when the very last dot-product is completed for an output feature.
Otherwise, the result of the dot-product continues to be stored in the accumulator.

The accumulators are implemented as shift-registers.
At any given cycle, each shift-register location contains the partial sum that belongs to a specific output feature.
The size of this shift-register depends on the latency $L$ of the dot-product unit.
That is, the same shift-register value that is used as the {\em init} value in the dot-product will be updated with the result of the dot-product, $L$ cycles later.
Hence, at any given cycle, each PE keeps $L$ different partial sums that belong to $L$ different output features for each dot-product unit.
Because each dot-product unit is fully-pipelined, $L$ different output computations are {\em interleaved}.
That is, for $L$ consecutive cycles, input features and filter weights for different output features will be fed into a dot-product unit in a sequence. In our implementation, we interleave both in the $W$ ($L_w$) and $H$ ($L_h$) direction.

Filter weights are stored in PE caches implemented in on-chip RAMs.
Every cycle, $W_{vec}\times C_{vec}$ transformed filter weights are loaded from these caches and fed onto the dot-product units.
Hence, $W_{vec}\times C_{vec}$ caches, or memory banks, are used in order to get the necessary on-chip memory read bandwidth.
A single filter weight can be loaded from each cache every cycle.

Filter weights are stored in the caches before the corresponding convolution layer starts.
To avoid idle computation cycles, the DLA uses double-buffering and overlaps convolutions with the PE cache updates.
While filter weights are loaded from the caches for a particular convolution layer, filter weights for the next convolution layer are also prefetched onto the caches.

Every cycle, the $W_{vec}$ outputs of each PE are sent to the ReLU unit for the Winograd output transform as explained in Section~\ref{sec:arithmetic_optimizations}.

\subsection{Stream Buffers}
\label{sec:feature_buffer}

\begin{figure}
\centering
\begin{subfigure}[b]{0.3\textwidth}
\centering
    \includegraphics[scale=0.4]{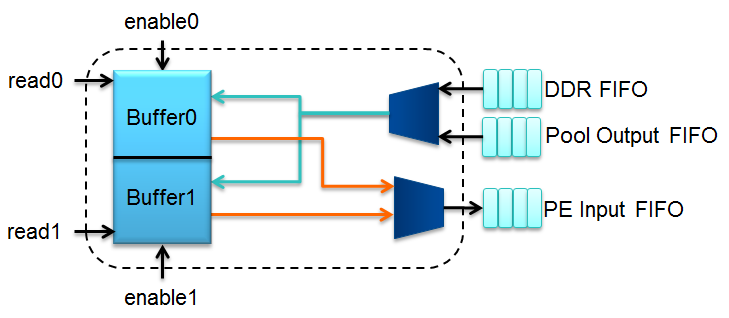}
    \caption{A single stream buffer.}
    \label{fig:stream_buffer_hardware}
\end{subfigure}
\\
\begin{subfigure}[b]{0.3\textwidth}
\centering
    \includegraphics[scale=0.4]{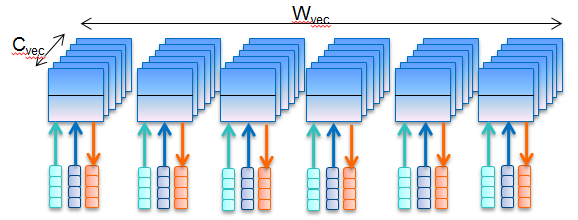}
    \caption{The array of stream buffers.}
    \label{fig:stream_buffer_array}
\end{subfigure}
\caption{Stream buffer hardware.}
\label{fig:stream_buffer}
\end{figure}

Stream buffers shown in Figure~\ref{fig:stream_buffer} are implemented in on-chip RAMs in order to store the feature data and to stream it to PEs.
Each stream buffer is double-buffered similar to filter caches.
Before the first convolution layer starts, the images are loaded from the DDR and stored in buffers. 
During the convolution layer execution, while feature data for a convolution layer is being streamed into the PEs, the outputs of convolutions are simultaneously stored into the buffers.
There are a total of $W_{vec}\times C_{vec}$ stream buffers.
The width of feature maps is divided into $W_{vec}$ buffers, and the depth is divided into $C_{vec}$ buffers.
Hence, a total of $W_{vec}\times C_{vec}$ stream buffers provide sufficient on-chip bandwidth for streaming an input feature region of size $1\times W_{vec}\times C_{vec}$ to PEs every cycle.
The output features, on the other hand, are generated in a different layout.
More specifically, each PE generates $Q_{vec}$ features in a cycle, hence, a total region of $1\times Q_{vec}\times K_{vec}$ is generated in a cycle.
A crossbar network is generated in order to store this region in $1\times W_{vec}\times C_{vec}$ buffers.

%
%
%

\subsection{Shared Exponent FP16}
\label{sec:shared_exp_fp16_compute}
Using half-precision (FP16) instead of single-precision (FP32) floating point operations can significantly reduce the resource requirement of each PE. However, although FP32 is natively supported on Arria 10's DSP blocks, FP16 is not, which leads to additional logic use.
To reduce this overhead, we use a shared exponent technique which allows us to perform the multiplications in fixed-point, which significantly reduces the overhead required to perform the FP16 dot-products. This technique works by leveraging the fact that Arria 10 DSP blocks can be fractured into two $18\times 18$ integer multipliers~\cite{AlteraA10} such that before sending the feature and filter data into each PE, we transform all
the values into 18-bit numbers, using the maximum exponent found in the group. Since the exponent matches for all the numbers, they can be treated as fixed-point numbers and can be sent directly to the $18 \times 18$ integer multipliers. After the dot-product is performed, the number is shifted back to 10-bits, and the exponent and sign bit is added back to the top 6 bits, reforming the 16-bit floating point value, which is stored back into the stream buffer. Note that the shared exponent transform is performed on the data prior to entering the PEs and thus only need to be applied once and can be shared across all PEs.

\subsection{Fully Connected Layers}
\label{sec:fully_connected_layers}

The DLA executes the fully-connected layers on the same PEs described in section~\ref{sec:pes}.
This approach makes the most efficient use of the dot-product units since these units are kept busy during the convolution and the fully-connected layers.

Due to the different characteristics of computations in fully-connected and convolutional layers, PEs need to be configured differently.
Specifically, the ratio of the total filter weights used in computations to the total amount of the computation is significantly higher in fully-connected layers than in convolution layers.
In other words, there is significantly less re-use of the fully-connected layer filter weights during the classification of a single image.
Hence, storing these filters in PE caches does not give any benefits.
Moreover, loading these filters from DDR uses significantly more bandwidth, which may become a performance bottleneck.


In order to alleviate the above issues, the DLA processes fully-connected layers in image batches.
After all convolution layers are completed layer by layer for a single image, the last layer will dump the image back out to external memory to batch up the images. Once a large enough batch of images are available, the batch of images is processed together during each of the fully-connected layers.
This allows sharing the fully-connected filter weights between the classification of different images, and hence, reducing the external memory bandwidth usage.
In other words, filters are shared and same filter weights are multiplied with different image features to produce the output features of different images.
This is in contrast to the convolution layers where features are shared and same features are multiplied with different filter weights to produce different output feature maps.
Hence, during the fully-connected layers, filter weights are streamed into the PEs and PE caches store the features for different images that are pre-loaded before computation starts.
The caches are sized to accommodate not only the convolution filters but also the batches of images that need to be processed in parallel during fully-connected layers.

The fully-connected layers are executed with the following configuration (summarized in Table~\ref{tbl:pe_configs}).
{\small
\begin{enumerate}
\item
No Winograd transformations are applied because features and filters are convolved to generate only a single output.
\item
Before starting the compute, features are pre-loaded into the PE caches.
For instance, if the image batch size is $S_{batch}$, each PE will store $N$ different image features, where $N$ is equal to $S_{batch}$/$K_{vec}$.
\item
During fully-connected layer computation, the $W_{vec}$/$N$ dot-product units in each PE are used to process one image.
\item
Each cycle, $F$ unique filter weights are loaded from the DDR and streamed into the PEs, where $F$ is equal to ($W_{vec}$/$N$)$\times$$C_{vec}$.
Each PE receives the same filter weights and multiplies them with different image features.
\item
Similar to the convolution configuration, $L$ different output computations are interleaved.
\item 
$W_{vec}$/$N$ partial sums in each PE are summed to produce $N$ outputs from each PE.
\end{enumerate}
}

\begin{table}[!h]
\centering
\begin{tabular}{|l|c|c|}
\hline
Configuration & Convolution & Fully-Connected \\
\hline
Winograd Transformation   & Yes & No \\
Batch Size &        1 & $S_{batch}$ \\
Streamed Data & Features & Filters \\
Cached Data & Filters & Features \\
Dot-Products per Image & $W_{vec}$ & $W_{vec}$/$N$ \\
\hline
\end{tabular}
\caption{Configuration of PEs during convolution and fully-connected layers.}
\label{tbl:pe_configs}
\end{table}

\subsection{Overall Architecture}
\label{sec:overall_architecture}

CNN algorithms often include other layers in addition to convolution and fully-connected layers.
For instance, AlexNet contains normalization, max-pooling, and ReLU layers.
Hence, our DLA contains additional hardware to support these different
types of layers to
enable the entire topology to be executed on the FPGA.

\begin{figure}
\centering
\includegraphics[scale=0.38]{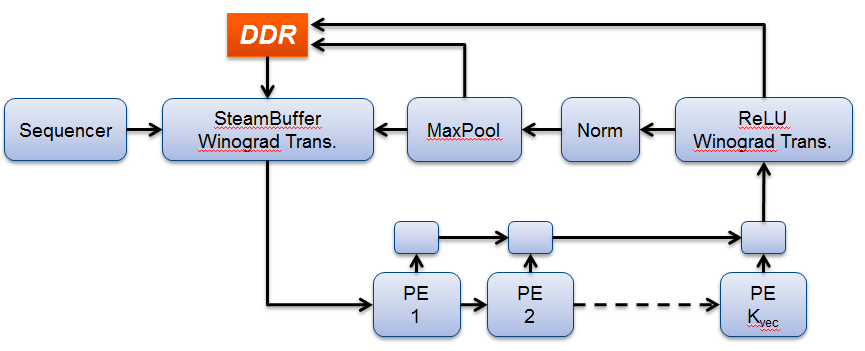}
\caption{Overall DLA Architecture.}
\label{fig:overall_architecture}
\end{figure}

Figure~\ref{fig:overall_architecture} shows the DLA hardware support for all the AlexNet layers.
The PEs, as discussed earlier, perform the dot-products for convolution and fully-connected layers.
The StreamBuffer unit manages the stream buffers, applies the Winograd transformations to features, and streams the transformed features to the first PE. The StreamBuffer unit also fetches the filter data from DDR and sends it to the PEs.
The features are forwarded through all the PEs via the daisy-chained input connections between them.
The outputs of the PEs are sent to the ReLU unit again via daisy-chained output connections.
ReLU unit applies the Winograd output transformations and non-linearity functions.
The throughput of ReLU unit and all the subsequent units are higher than the total throughput of the PEs in order to avoid stalls.
The outputs of the ReLU unit are sent to the normalization unit, which applies the normalization formula across the feature maps.
Because PEs compute feature maps in tiles, normalizing a tile requires buffering of convolution outputs from the previous tile.
The outputs of the normalization unit are sent to the pooling unit which computes the maximum value in a window.
Because each feature map is pooled independently, no data buffering is necessary between the feature map tiles.
If more convolution layers are to follow, the output of the pooling unit is stored back onto the stream buffer for further processing.
Following the last convolution layer, the outputs of the pooling unit are stored to external memory.
At the start of the fully-connected layers, these features are read back from external memory and loaded onto the PE caches as described earlier.
Also, the ReLU output is sent directly to DDR, without applying Norm or Pool.

The DLA includes several control signals, such as stream buffer read/write addresses, PE cache read/write addresses, PE done/reset signals, normalization/pooling bypass signals, etc.
These signals are generated by the sequencer unit which is configured according to the topology of the CNN algorithm being executed.
The sizes of input/output images, intermediate features, filters, normalization/pooling windows, convolution strides, etc. are used in calculating the exact cycles certain actions are taken at or the addresses that are accessed.
Hence, executing a different CNN algorithm on the same hardware requires just changing the sequencer configuration.

In AlexNet, the normalization and pooling operations are not always performed after every convolution layer.
Hence, these units can be by-passed depending on the topology that is being executed. This makes extending this architecture to support software configurability relatively straightforward because the sequence of layers bypassed can be changed after the FPGA is programmed and is similar to the work in~\cite{Zhang2016}.

All the units described above are written as OpenCL kernels.
Each kernel executes independently and concurrently.
The connections between the kernels are implemented as FIFOs using the Intel channel API.

\section{Design Space Exploration and Analytical Models}
\label{sec:performance_modeling}

One of the benefits of our architecture is that the resource
usage of the PE array, stream buffers, and filter caches
can be analytically modeled using the $C_{vec}$, $K_{vec}$, $W_{vec}$, and $Q_{vec}$ parameters.

For 16-bit floating point precision, equation~\ref{equ:dsps} models the DSP usage for all the PE elements, which assumes that each DSP block can perform two 16-bit floating point multiplies and two 16-bit floating point adds and no Winograd.  If Winograd is applied, we divide equation~\ref{equ:dsps} by 2 and add on a constant factor of 200. The constant factor is an over estimate, and accounts for the on-chip Winograd transforms as shown in Figure~\ref{fig:winograd_flow} and the value chosen is applicable to the $F(4,3)$ Winograd transforms we use. The stream buffers and filter
caches M20K usage can be modeled using equation~\ref{equ:m20k_st_buf} and~\ref{equ:m20k_cache} respectively, which assumes a given M20K can store 1024 16-bit floating point values by forming a 2 word wide by 512 deep memory~\cite{AlteraA10}. Equation~\ref{equ:m20k_st_buf} models the number of M20Ks required to store the largest input and output feature map for any given layer (represented by the $MAX( Depth_{in} + Depth_{out} )$). $C$ is the number of input feature maps for a given layer, $H$ and $W$ are the feature map height and width. Equation~\ref{equ:m20k_cache} models the number of M20Ks required to store all filter weights for a single output feature map. This is scaled up by the number of PEs (i.e. $K_{vec}$) since each PE processes one output feature map at any given time. Note that for the filter caches the depth is not considered since the filters don't require the entire M20K depth (i.e. there are less than 512 words needed).
\begin{equation} \label{equ:dsps}
    N_{dsps} = (W_{vec}-Q_{vec}+1) \times  Q_{vec} \times  K_{vec} \times  C_{vec} \times  0.5
\end{equation}
\begin{equation} \label{equ:m20k_st_buf}
    \begin{split}
	N_{banks} &= W_{vec}\times C_{vec} \\
	Depth &= C \times  W \times  H / N_{banks} \\
	N_{M20K} &= CEIL( \frac {MAX( Depth_{in} + Depth_{out} )}{ 512 \times  2 } ) \times  N_{banks}
\end{split}
\end{equation}
\begin{equation} \label{equ:m20k_cache}
    \begin{split}
	N_{banks} &= W_{vec}\times C_{vec} \\
	N_{M20K} &=  N_{banks} \times  K_{vec} / 2
\end{split}
\end{equation}

Expected throughput is modeled using the vector dimensions ($C_{vec}$, $K_{vec}$,$W_{vec}$, and $Q_{vec}$),
feature map sizes, output map sizes, filter sizes, and DDR bandwidth utilization. 
For a single convolutional layer, the number of cycles to process an image is shown in
equation~\ref{equ:conv_latency}. Here,  $C$ is the number of input feature maps, $K$ is the number of
output feature maps, $Q$ is the width of the output feature map, $P$ is the height of
the output feature map.
$DSP_{eff}$ represents the efficiency of the DSPs and models any quantization issues due to $Q_{vec}$ and
interleaving width-wise ($L_w$) and height-wise ($L_h$) as described in
section~\ref{sec:pes}, where we ignore quantization effects on $C$
and $K$ for simplicity (e.g. if $C_{vec}$ does not divide $C$ evenly). For example, if the output image is 20 wide, and
$Q_{vec}=3$ with no interleaving (i.e. $L_h=L_w=1$), on the 7th cycle, only the first two values of $Q_{vec}$
will have useful output, the last value will be dropped. In this case, the
$DSP_{eff}=20/(CEIL(6.67)\times 3) = 95\%$. 
$N_{cycles}$ are the number of cycles required to generate all output
feature-maps for the layer, assuming there are no memory bandwidth
constraints.
$BYTE_{req}$ are the total bytes of prefetched filter weights required to be loaded
from DDR during the convolution layer where $R_{next}$ and $S_{next}$ are the filter dimensions of the next convolution layer, and $C_{next}$ is the number of feature map layers in the next convolutional layer,
and $BYTE_{ddr}$ are the total number of bytes that can be transferred during the convolutional layer
assuming  that there is one DDR memory interface that is 64 bytes wide.
$N_{real}$ is the estimated number of cycles required taking into
account any DDR bandwidth limitations of the device.
\begin{equation} \label{equ:conv_latency}
    \begin{split}
	DSP_{eff} &= Q / ( CEIL(Q/(Q_{vec}\times L_w))\times Q_{vec}\times L_w) \times  \\
                         &  P / ( CEIL(P/(L_h))\times L_h) \\
    N_{flops} &= 2 \times  K \times  C \times  Q \times  P \times  DSP_{eff} \\
    N_{cycles} &= N_{flops} / ( N_{dsps} \times  2 ) \\
	BYTE_{req} &= K_{next} \times  R_{next} \times  S_{next} \times  C_{next} \times  2\\
    BYTE_{ddr} &= 64 \times  N_{cycles} \\
    N_{real} &= N_{cycles} \times  BYTE_{req} / BYTE_{ddr}
\end{split}
\end{equation}

For fully-connected layers, the number of cycles is shown in equation~\ref{equ:fc_latency} which calculates the cycles required for an entire batch of images. Here, $K$ and $C$ are the number of input and output feature maps for the fully-connected layer and $S_{batch}$ is the batch size used. For fully-connected layers, $BYTE_{req}$ are the total bytes of the filter weights that need to be loaded. Also, we ignore any quantization effects for fully-connected layers, since empirically we show that DSP efficiency is close to 100\% (shown later in Table~\ref{tab:gflops}).
\begin{equation} \label{equ:fc_latency}
    \begin{split}
	S_{batch} &= K_{vec}\times 2 \\
    N_{flops} &= 2 \times  K \times  C \times  S_{batch} \\
    N_{cycles} &= N_{flops} / ( N_{dsps} \times  2 ) \\
    BYTE_{req} &= C \times  K \times  2\\
    BYTE_{ddr} &= 64 \times  N_{cycles} \\
    N_{real} &= N_{cycles} \times  BYTE_{req} / BYTE_{ddr}
\end{split}
\end{equation}

To get the final throughput in terms of images per second, we divide the
clock frequency of the design by the total cycles for all layers to
process.  For fully-connected layers, we have to normalize to one
image so we divide by the batch size, $S_{batch}$. We ignore the
execution time of other layers, such as Norm and ReLU, since these are
executed concurrently with the convolutional or fully-connected layers,
and have a negligible execution overhead.
\begin{equation} \label{equ:fc_latency}
    T_{all} = f_{max} / ( \Sigma_{conv}( N_{real} ) + \Sigma_{fc}( N_{real} / S_{batch} ) ) 
\end{equation}

Using both the resource usage estimates and throughput models,
we can find the optimal $C_{vec}$ and $K_{vec}$ value for a given FPGA device, assuming all other values are set by the user (e.g. $f_{max}$, $W_{vec}$, etc).
A curve of this is shown in the results section in Figure~\ref{fig:plot_throughput}.

\section{Experimental Evaluation}
\label{sec:experimental_eval}

We evaluate our DLA by implementing the AlexNet topology on Intel's Arria 10 dev kit which contains
a A10-1150 device (20nm). We use a batch size of 1 for convolution layers, and 96 for the fully connected layers as described in Section~\ref{sec:fully_connected_layers}. We use only one bank of DDR4x64 at 1200MHz with a total bandwidth of 17GB/s to reduce the power required for the FPGA.
We compare against the
work in~\cite{Suda2016} and~\cite{Zhang2016}. Additionally, we compare
against the \emph{best known} results for nVidia's TitanX GPU (28nm) taken from~\cite{ConvNet}.
Note that nVidia used 28nm for its last generation GPU and skipped the 20nm node, which is why
TitanX is used in this comparison. When measuring throughput, we measure the total system throughput, which includes all the data transfers of the images to the FPGA using the ILSVRC data set~\cite{ILSVRC},
which would be incurred in a real application, which is not done
in~\cite{Zhang2016} nor~\cite{ConvNet}. In order to hide the
latency of the transfers, we pipeline the execution of the DLA with the image data transfers from host to FPGA DDR memory.
Also note that the data precision vary from fixed and floating point in the studies in~\cite{Suda2016,Zhang2016,ConvNet} and our work. Previous work~\cite{Suda2016,icml2015_gupta15} have shown the limited impact of 16-bit fixed point when compared to 16-bit floating point and is not described here.

\section{Results}
\label{sec:results}
To illustrate the efficiency of our architecture, we show the GFLOPS of
the DLA for each fully-connected and convolutional layer in
Table~\ref{tab:gflops} as well as the DSP efficiency.  Here, we define DSP
efficiency as the percentage of time the DSP is occupied and doing
useful computation.
\begin{table}
\centering
\begin{tabular}{|l|c|c|c|}
    \hline
  Layer & Eff. GFLOPS & Act. GFLOPS & Eff. \\
    \hline
    Conv1 & 2,308 & 1,154 & 82.9\% \\
    Conv2 & 1,740 & 870 & 62.5\% \\
    Conv3 & 1,960 & 980 & 72.4\% \\
    Conv4 & 1,960 & 980 & 72.4\% \\
    Conv5 & 1,743 & 871 & 62.6\% \\
    Fc6 & 1,389 & 1,389& 99.8\% \\
    Fc7 & 1,386 & 1,386 & 99.6\% \\
    Fc8 & 1,378 & 1,378& 99.0\% \\
    \hline
\end{tabular}
\caption{The average GFLOPS achieved of convolutional and fully-connected layers and DSP efficiency when using an $8\times 48$ configuration. Shows both effective GFLOPS (Eff. GFLOPS) due to Winograd and actual GFLOPS (Act. GFLOPS).}
\label{tab:gflops}
\end{table}

It is clear that for most layers, DSP efficiency and GFLOPS are
relatively high, which is required to be competitive against the GPU.
The DSP efficiency differs between layers because vectorization factors
($W_{vec}$, $Q_{vec}$, $K_{vec}$, $C_{vec}$, $S_{vec}$) lead to different quantization inefficiencies for
different feature, filter and output dimensions as described in
equation~\ref{equ:conv_latency}. For instance, Conv2 has
the lowest efficiency because it uses $5\times 5$ filter weights which are
sub-optimally vectorized with $1\times 3$ tile sizes used. Moreover, FC layers have
close to ideal efficiency because the dimensions of their input
features, filter weights, and output features are large with respect to
the vectorization factors, i.e. how many features and filters are loaded
and how many output features are computed every cycle as discussed in
Section~\ref{sec:fully_connected_layers}.

We should note that for Conv1, we achieve a high efficiency even
though the number of input feature maps for the first layer is three, which 
is not
wide enough to fill up the vector 8 dot-product units in each PE (i.e. 3 is less than $C_{vec}=8$).
In order to get around this limitation,
we fold the three input feature maps to create 48 sub-feature maps, such that we can
saturate the dot-product width. 

\begin{figure}
\centering
\includegraphics[scale=0.4]{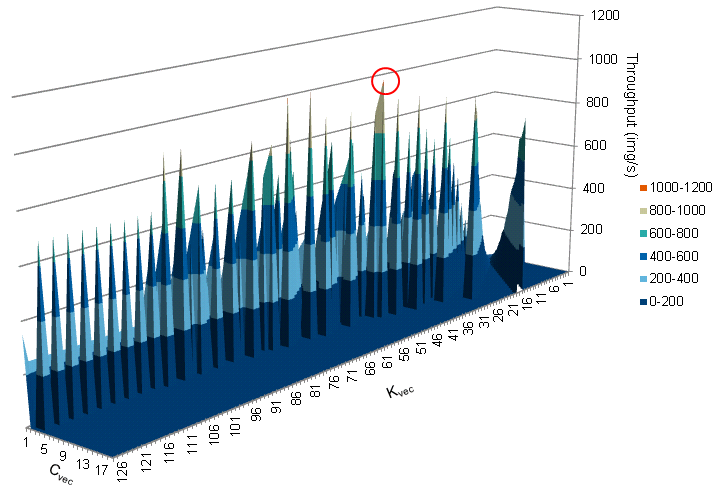}
\caption{Plot of expected throughput for various $C_{vec}$ and $K_{vec}$ values.}
\label{fig:plot_throughput}
\end{figure}

Figure~\ref{fig:plot_throughput}  plots the achievable throughput for various $C_{vec}$ and $K_{vec}$ values, using the Arria 10 1150 device. Here, we assume $f_{max}$ is 300MHz, $Q_{vec}=4$, and $W_{vec}=6$, and we only explore positions where $K_{vec}$ are even multiples of $C_{vec}$ (areas which are not even multiples are 0 in Figure~\ref{fig:plot_throughput}), which leads to a more efficient memory structure for the stream buffers and filter cache. Note that the highlighted red circle is one of the peak throughput numbers with $C_{vec}=8$ and $K_{vec}=48$. This is our final configuration which achieves a throughput of 1020 { \small img/s}.

\begin{figure}
\centering
\includegraphics[scale=0.4]{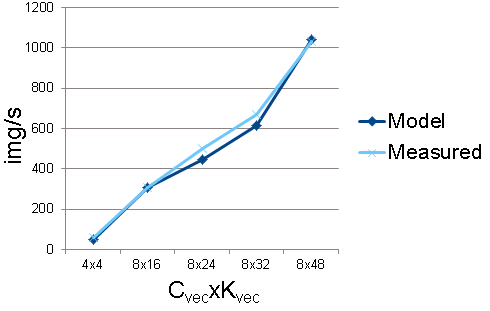}
\caption{A comparison of empirical data against analytical model for
A10-1150 device.}
\label{fig:model}
\end{figure}

To validate our analytical models presented in
Section~\ref{sec:performance_modeling}, we plot predicted { \small img/s} given by
our models and the measured performance, as shown in
Figure~\ref{fig:model}. Note that in Figure~\ref{fig:model} we scale down the model { \small img/s} predictions provided by equations~\ref{equ:conv_latency} and~\ref{equ:fc_latency} by 16\% to
account for any inefficiencies in the pipelined data transfers and the overhead of data movement between the host processor and FPGA, which is included in the measured throughput values. 16\% is used because this was measured as the average difference between the system-level throughput and the FPGA device throughput.
As shown in the graph, our model throughput predictions match very closely to the actual measurements.

\subsection{Resource Usage}

To show the impact of the shared exponent floating point optimization described in Section~\ref{sec:shared_exp_fp16_compute}, we show the
resource usage of a single PE using true half-precision
dot-products
(Half-type) vs shared exponent dot-products in Table~\ref{tab:fpga_res_fp16}. The shared exponent significantly reduces resource usage since we can leverage the DSP fully, whereas when using the half-type, a lot of logic must be used to normalize and compute the 16-bit floating point multiplications and perform the dot-product. Also, it should be noted that no impact to accuracy was seen to the top-1 and top-5 error rate (56\% and 79\% respectively) between our shared exponent implementation and 32-bit floating point.
\begin{table}
\centering
    \begin{tabular}{|c|c|c|c|c|}
	\hline
	FP16 config & ALMs & Reg \\ \hline
	Half-type & 10.7K & 26K \\
	Shared Exponent & 3.3K & 10.6K \\
	\hline
    \end{tabular}
\caption{Resource usage of PE without shared exponent optimizations (Half-type) and with shared exponent optimizations.}
\label{tab:fpga_res_fp16}
\end{table}

Table~\ref{tab:fpga_res} shows the final resource
usage of an $8\times 48$ ($C_{vec}\times K_{vec}$) configuration running on the
Arria 10 1150 device running at 303MHz.
\begin{table}
\centering
    \begin{tabular}{|c|c|c|c|c|c|}
	\hline
	ALMs & Reg & {\small M20K } & DSPs & Freq. \\ \hline
	{\small 246K (58\%) } &	{\small 681K } & { \small 2487 (92\%) } &	{\small 1476 (97\%) } & {\small 303 MHz } \\
	\hline
    \end{tabular}
\caption{Resource usage and clock frequency on the Arria 10 1150 device, for an $8\times 48$ configuration running at 303MHz.}
\label{tab:fpga_res}
\end{table}

\subsection{FPGA Comparisons to the state-of-the-art}
Table~\ref{tab:fpga_comp} and Table~\ref{tab:gpu_comp} shows our comparisons
against prior work on FPGAs and GPUs respectively. As
Table~\ref{tab:fpga_comp} shows, we achieve 8.4x more GFLOPS
when compared to the latest Ultrascale (KU 20nm~\cite{KU}) result, which uses a batch size 32 for the fully-connected layers, and 19x more GFLOPS than the latest Stratix V result, both running AlexNet.
It is important to note that in~\cite{Zhang2016} the authors are only
able to use 50\% of DSP
resources and claim that this is due to a limitation in SDAccel when
using partial reconfiguration. However, even if they were able to use
100\% of DSPs, the 8.4x gap would still not be closed
since they are only able to
achieve a 14.7\% efficiency of their DSPs, which assumes a 1.1 TOPS
for the KU060 device at 200MHz used in~\cite{Zhang2016}.
\begin{table}
\centering
    \begin{tabular}{|c|c|c|}
	\hline
	Stratix V (28nm)\cite{Suda2016} & KU060 (20nm) \cite{Zhang2016} & DLA (20nm) \\
	\hline
	72.4 GOPS & 165 GOPS & 1382 GFLOPS  \\
	\hline
    \end{tabular}
\caption{A comparison of our DLA against~\cite{Zhang2016,Suda2016} for AlexNet. For
the DLA, effective flops shown due to Winograd.}
\label{tab:fpga_comp}
\end{table}

We should note that in~\cite{Zhang2016} and~\cite{Suda2016}, they show better GOPS
numbers for VGG of 266 GOPS and 118 GOPS respectively. Since our
architecture is also applicable to VGG, which is based on convolutional
and fully-connected layers as well, our performance will not
be impacted negatively with the VGG topology. In fact, since VGG is more regular, DSP efficiency is improved on previous work~\cite{Zhang2016} and we believe that this should also benefit the DLA architecture.

Finally, in Table~\ref{tab:gpu_comp} we show a comparison of our work against
the best known nVidia results for the TitanX and M4 running
AlexNet with image sizes of $224\times 224$ and batch size 128 (note that the M4 white paper doesn't specify batch size) . The TitanX card has a peak 6.1 TFLOPS, compared to the 1150 Arria
10 device which has 1.3 TFLOPS and as Table~\ref{tab:gpu_comp} shows, the
TitanX is able to beat our work in terms of raw performance.  However,
when normalized against power consumption, we are competitive
with both
nVidia devices~\cite{ConvNet,M4Nvidia}. 
Also note
that the { \small img/s}/W numbers shown in Table~\ref{tab:gpu_comp} are 5.8x better than the { \small img/s}/W for
AlexNet presented in~\cite{Zhang2016}.
\begin{table}
\centering
    \begin{tabular}{|c|c|c|c|c|}
	\hline
	& { \tiny img/s} & {\tiny Watts (W brd) } & {\tiny Peak Ops } & { \tiny img/s/W} \\
	\hline
	DLA (20nm) & 1020 & 45 & 1.3{\tiny TFLOPS} & 23 \\
	KU060 (20nm) & 104 & 25 & 3.6{\tiny TOPS} & 4 \\
	TitanX (28nm) & 5120 & 227 & 6.1{\tiny TFLOPS}& 23 \\
	M4 (28nm) & 1150 & 58 & 2.2{\tiny TFLOPS} & 20 \\\hline
    \end{tabular}
    \caption{A comparison of the DLA at 303MHz against~\cite{Zhang2016} and \cite{ConvNet,M4Nvidia}. KU060 peak operations are integer, the rest are 32-bit floating pt.}
\label{tab:gpu_comp}
\end{table}

\subsubsection{Discussion on performance comparisons}
There are several simplifications the authors do in~\cite{ConvNet} which can significantly boost the performance of the TitanX result shown in Table~\ref{tab:gpu_comp} including the \emph{removal} of communication overhead and the use of random data instead of the ILSVRC database set. As such, we suspect that the raw performance numbers are overly optimistic and, unlike our throughput measurements, do not reflect the actual throughput of a production system.
Additionally, the KU060 104 { \small img/s} is estimated using Figure 10d in~\cite{Zhang2016}, which assumes no execution overhead for data transfers and ignores the execution time of the non-linear layers (i.e. Pool, Norm, ReLU), which again is overly optimistic. Due to these simplifications, we suspect that the relative system performance benefit of our DLA is much larger than what is reported in Table~\ref{tab:gpu_comp}.

\section{Conclusions}
\label{sec:conclusion}


We describe a novel architecture written in OpenCL, DLA, targeted for computing CNNs on
FPGAs. We demonstrate an approach that reduces the required memory
bandwidth by an order-of-magnitude through the use of an on-chip stream
buffer that efficiently stores input and output feature maps.
Additionally, we demonstrate a vectorization approach that achieves over
60\% DSP efficiency and uses the Winograd transform to significantly
reduce the DSPs required to perform the convolution layers. Because of
these improvements, we are able to achieve an overall system-level
performance that is 10x faster than the state-of-the-art on FPGAs when
running AlexNet and is competitive in energy efficiency with the best
known results on nVidia's TitanX GPU at 23 {\small img/s}/W.

Future work includes mapping other CNNs such as GoogLeNet and VGG to our
architecture, and exploring how run-time reconfigurability may impact
performance of our architecture.

%
\section{Acknowledgements}
\label{sec:acknowledgements}
We would like to thank Stephen Weston for his insightful comments and
Kevin Jin for the experimental data.

%
\bibliographystyle{abbrv}

{ \small
\bibliography{cnn}  

\begin{thebibliography}{10}

\bibitem{AlteraA10}
Altera.
\newblock {\em {Arria 10 Device Overview}}.
\newblock White Paper A10-OVERVIEW. Altera Corporation, Jan. 2015.

\bibitem{Cadambi2010}
S.~Cadambi, A.~Majumdar, M.~Becchi, S.~Chakradhar, and H.~P. Graf.
\newblock A programmable parallel accelerator for learning and classification.
\newblock In {\em Proceedings of the 19th International Conference on Parallel
  Architectures and Compilation Techniques}, PACT '10, pages 273--284, New
  York, NY, USA, 2010. ACM.

\bibitem{ConvNet}
S.~Chintala.
\newblock {convnet-benchmarks}, 2016.

\bibitem{Farabet2009}
C.~Farabet, C.~Poulet, J.~Y. Han, and Y.~LeCun.
\newblock Cnp: An fpga-based processor for convolutional networks.
\newblock In {\em 2009 International Conference on Field Programmable Logic and
  Applications}, pages 32--37, Aug 2009.

\bibitem{icml2015_gupta15}
S.~Gupta, A.~Agrawal, K.~Gopalakrishnan, and P.~Narayanan.
\newblock Deep learning with limited numerical precision.
\newblock In D.~Blei and F.~Bach, editors, {\em Proceedings of the 32nd
  International Conference on Machine Learning (ICML-15)}, pages 1737--1746.
  JMLR Workshop and Conference Proceedings, 2015.

\bibitem{OpenCLKhronos}
Khronos.
\newblock {The open standard for parallel programming of heterogeneous
  systems}, 2015.

\bibitem{AlexNet}
A.~Krizhevsky, I.~Sutskever, and G.~E. Hinton.
\newblock Imagenet classification with deep convolutional neural networks.
\newblock In {\em Advances in Neural Information Processing Systems}, NIPS '12,
  pages 1097--1105, 2012.

\bibitem{DBLP:journals/corr/Lavin15b}
A.~Lavin.
\newblock Fast algorithms for convolutional neural networks.
\newblock {\em CoRR}, abs/1509.09308, 2015.

\bibitem{NvidiaTitanX}
{nVidia}.
\newblock {GPU-Based Deep Learning Inference: A Performance and Power
  Analysis}, November 2015.

\bibitem{M4Nvidia}
{NVIDIA}.
\newblock Nvidia(r) tesla(r) m4 gpu accelerator, Apr. 2016.

\bibitem{Microsoft}
K.~Ovtcharov, O.~Ruwase, J.-Y. Kim, J.~Fowers, K.~Strauss, and E.~Chung.
\newblock Accelerating deep convolutional neural networks using specialized
  hardware, February 2015.

\bibitem{Peemen2013}
M.~Peemen, A.~A.~A. Setio, B.~Mesman, and H.~Corporaal.
\newblock Memory-centric accelerator design for convolutional neural networks.
\newblock In {\em 2013 IEEE 31st International Conference on Computer Design
  (ICCD)}, pages 13--19, Oct 2013.

\bibitem{Qiu2016}
J.~Qiu, J.~Wang, S.~Yao, K.~Guo, B.~Li, E.~Zhou, J.~Yu, T.~Tang, N.~Xu,
  S.~Song, Y.~Wang, and H.~Yang.
\newblock Going deeper with embedded fpga platform for convolutional neural
  network.
\newblock In {\em Proceedings of the 2016 ACM/SIGDA International Symposium on
  Field-Programmable Gate Arrays}, FPGA '16, pages 26--35, New York, NY, USA,
  2016. ACM.

\bibitem{Sankaradas2009}
M.~Sankaradas, V.~Jakkula, S.~Cadambi, S.~Chakradhar, I.~Durdanovic,
  E.~Cosatto, and H.~P. Graf.
\newblock A massively parallel coprocessor for convolutional neural networks.
\newblock In {\em Proceedings of the 2009 20th IEEE International Conference on
  Application-specific Systems, Architectures and Processors}, ASAP '09, pages
  53--60, Washington, DC, USA, 2009. IEEE Computer Society.

\bibitem{ILSVRC}
{Stanford Vision Lab}.
\newblock Imagenet large scale visual recognition challenge (ilsvrc), 2015.

\bibitem{Suda2016}
N.~Suda, V.~Chandra, G.~Dasika, A.~Mohanty, Y.~Ma, S.~Vrudhula, J.-s. Seo, and
  Y.~Cao.
\newblock Throughput-optimized opencl-based fpga accelerator for large-scale
  convolutional neural networks.
\newblock In {\em Proceedings of the 2016 ACM/SIGDA International Symposium on
  Field-Programmable Gate Arrays}, FPGA '16, pages 16--25, New York, NY, USA,
  2016. ACM.

\bibitem{DBLP:journals/corr/SzegedyLJSRAEVR14}
C.~Szegedy, W.~Liu, Y.~Jia, P.~Sermanet, S.~E. Reed, D.~Anguelov, D.~Erhan,
  V.~Vanhoucke, and A.~Rabinovich.
\newblock Going deeper with convolutions.
\newblock {\em CoRR}, abs/1409.4842, 2014.

\bibitem{Winograd}
S.~Winograd.
\newblock {\em Arithmetic Complexity of Computations}, volume~33.
\newblock Siam, 1980.

\bibitem{KU}
Xilinx.
\newblock {\em {UltraScale Architecture and Product Overview}}.
\newblock Preliminary Product Specification. Xilinx Corporation, June 2016.

\bibitem{Zhang2016}
C.~Zhang, Z.~Fang, P.~Zhou, and J.~Cong.
\newblock Caffeine: Towards uniformed representation and acceleration for deep
  convolutional neural networks.
\newblock In {\em Proceedings of the 2016 International Conference On Computer
  Aided Design}, ICCAD '16, New York, NY, USA, 2016. ACM.

\bibitem{Zhang2015}
C.~Zhang, P.~Li, G.~Sun, Y.~Guan, B.~Xiao, and J.~Cong.
\newblock Optimizing fpga-based accelerator design for deep convolutional
  neural networks.
\newblock In {\em Proceedings of the 2015 ACM/SIGDA International Symposium on
  Field-Programmable Gate Arrays}, FPGA '15, pages 161--170, New York, NY, USA,
  2015. ACM.

\end{thebibliography}
}
\end{document}